\documentclass[12pt]{article}
\usepackage{amsmath,amssymb,amsfonts,amsbsy}
 \usepackage{setspace}
\usepackage{graphicx}
\newcommand{\ba}{\begin{eqnarray}}
\newcommand{\ea}{\end{eqnarray}}
\newcommand{\beqs}{\begin{eqnarray}}
\newcommand{\eeqs}{\end{eqnarray}}

\begin{document}
%%%%%%% please do not touch these! %%%%%%
\setcounter{section}{0}
\setcounter{subsection}{0}
\setcounter{equation}{0}
\setcounter{figure}{0}
\setcounter{footnote}{0}
\setcounter{table}{0}
\begin{center}
\textbf{FLAVOR DEPENDENCE OF THE SPIN-INDEPENDENT AND SPIN-DEPENDENT
PARTS OF GPDs(x,t,$\xi=0$)
 } %  \bigskip

\vspace{5mm}

 \underline{O.V. Selyugin}\footnote{selugin@theor.jinr.ru}
%}\bigskip

\vspace{5mm}

\begin{small}

%{\small
  $^1$\it{BLTPh, Joint Institute for Nuclear Research, Dubna, Russia} \\
%$^{\,2}$\it IFPA, AGO Dept., Universit\'e de Li\`ege, Li\`ege, Belgium\\
%$^{\,3}$ \it Dipartamento di Fisica Teorica, Universit\`a di Torino, Italy\\
%}
\end{small}
\end{center}

\vspace{0.0mm} % Don't laugh: it does change the spacing!

\begin{abstract}
 The different sets of  PDF with
the new form of t-dependence of generalized parton distributions (GPDs) were examined
in the descriptions  of the  electromagnetic form factors of the proton and neutron.
One of the purposes  was to minimize the number of  fitting parameters.
We found that main flavor difference related to the spin-dependent of PDF
incoming as part in  GPDs. Hence, contrary to some other work, our result shows
a little flavor dependence of the t-dependence of the $GPDs(x,t,\xi=0$).
\end{abstract}
\vspace{7.2mm}
%\section{Introduction}

   The parton picture of the hadron % structure
   is in most part represented by the parton distribution functions (PDFs).
   They are determined in the deep inelastic processes. The next step in the development of the picture of the hadron
% structure
 was made by introducing the non-forward structure functions - general parton distributions - GPDs
   \cite{Ji97}  with spin-independent the $H(x,\xi,t) $
   and the spin-dependent $E(x,\xi,t)$ parts.
     Generally,  GPDs depend on the momentum transfer $t$, the average momentum fraction
  $x=0.5 (x_i + x_f)$ of the active quark, and the skewness parameter
  $2 \xi=x_f - x_i$ that measures the longitudinal momentum transfer. Some of the advantages of  GPDs were presented by the sum rules \cite{Ji97}
   %which put the connections of the  GPDs with the standard electromagnetic hadron form factors.
  \ba
 F_{1}^q (t) = \int^{1}_{0} \ dx  \ {\cal{ H}}^{q} (x,\xi=0, t),  \ \ \
%\ea
%\ba
 F_{2}^q (t) = \int^{1}_{0} \ dx \  {\cal{E}}^{q} (x, \xi=0, t).
\ea

Now we cannot obtain the $t$-dependence of GPDs  from the first principles,
but it can be obtained from the phenomenological description by $GPDs$
of the nucleon electromagnetic form factors. Many different forms
  of the $t$-dependence of GPDs were proposed.
   In the quark diquark model \cite{Liuti1,Liuti2} the form of  GPDs
   consist of three parts - PDFs, function distribution and Regge-like.
\ba
F_{q}(x,t)=N_{q} \ G_{M_{x}^{I.II}}^{\lambda^{I.II}}(x,t) \ R_{Pq}^{\alpha_{q} \alpha_{q}^{\prime}}(x,t)
 \ea
  The parameters have the flower dependence for the all three parts.
  As  a result, they came to the conclusion:
"The data show, in particular, a suppression of $d$ quarks with respect to $u$ quarks at large
momentum transfer".
 In other works (see e.g. \cite{Kroll04})
  the description of the $t$-dependence of  GPDs  was developed
  in a  more complicated picture using the polynomial forms with respect to $x$.
    Note that  in \cite{Yuan03}  %,Burk04}
    it was shown that at  large $x  \rightarrow 1$
    and momentum transfer the behavior of GPDs
  requires a larger power of $(1-x)^{n}$ in the  $t$-dependent exponent:
\ba
{\cal{H}}^{q} (x,t) \  \sim  exp [ a \ (1-x)^n \ t ] \ q(x).
\ea
with $n \geq 2$. It was noted that $n=2$ naturally leads to the Drell-Yan-West duality
 between parton distributions at large $x$ and the form factors.

% \section{New momentum transfer dependence of GPDs }

Let us modify the original Gaussian ansatz
% in order to incorporate
% the observations of \cite{R98} and \cite{Burk04}
 and choose  the $t$-dependence of  GPDs in the simple form
\ba
{\cal{H}}^{q} (x,t) \  = q(x) \   exp [  a_{+}  \
    (1-x)^2/x^{m}  \ t ].
%\frac{(1-x)^2}{x^{m} } \ t ].
\ea
  The value of the parameter $m=0.4$ is fixed by the low $t$  experimental data while
 the free parameters $a_{\pm}$ ($a_{+} $ - for ${\cal{H}}$
and $a_{-} $ - for ${\cal{E}}$) were chosen to reproduce the
experimental data in the whole $t$ region.
  The isotopic invariance can be used to relate the proton and neutron GPDs.
 Hence, we do not change any parameter
 and keep the same $t$-dependence of GPDs as in the case of  proton.

In our first  work \cite{ST-PRDGPD} the function $q(x)$
 is based on the MRST02 global fit \cite{MRST02}.
 In all calculations we restrict ourselves %, as in other quoted works,
  to  the contributions of only valence  $u$ and $d$ quarks.
 Following the standard representation %, see for example \cite{R04},
  we have for the Pauli form factor $F_2$
\ba
 {\cal{E}}^{q} (x,t) \  = &&
{\cal{E}}^{q} (x) \   exp [  a_{-}  \  (1-x)^{2}/x^{0.4} \ t];    \\ \nonumber
% \ea
% with
% \ba
{\cal{E}}^{u} (x) \  = && k_{u}/N_{u} \  (1-x)^{\kappa_1} \ u(x), \ \ \
%\ea
% \ba
{\cal{E}}^{d} (x) \  = k_{d}/N_{d} \  (1-x)^{\kappa_2} \ d(x),
\ea
 where $\kappa_1 =1.53$ and $\kappa_2=0.31$   \cite{R04}.
According to  the normalization of the Sachs form factors, we have
$k_u=1.673  , \ \ \  k_d=-2.033,   \  \  \  \   \ N_u=1.53  , \ \ \  N_d=0.946   $

    Now  many PPDs, proposed by different Collaborations,  were examined
 %   with the view point - how they can describe the whole sets of the experimental data
         to compare the descriptions
    of the electromagnetic form factors of the proton and neutron. We take 464 experimental data
    and take into account only statistical errors. As a result, we find that the different
    PDF sets, which well describe the deep inelastic processes, gave the large difference
    in the description of the form factors \cite{Sel-Sp12}. The whole sets of the results will be published.
    Now we note that a better description of the form factors was given by
    PDFs of the \cite{ABKM09,ABM12} and  \cite{Kh12}.
                 The obtained description of the electromagnetic form factors is shown on Fig. 1 (left) for
     the proton and Fig.1(right) for the neutron. Note that at small momentum transfer practically all PDFs  gave the same descriptions. However, at large $t$ we obtain the different description for the different PDFs.

 %Finally, we present our calculations for the ratio of $G_{E}^{n}/G_{M}^{n}$
 %for neutron in Fig.1 (left).

   %==================Figs.1 =============================
\label{sec:figures}
\begin{figure}
\vspace{-1cm}
\begin{center}
\includegraphics[width=0.4\textwidth] {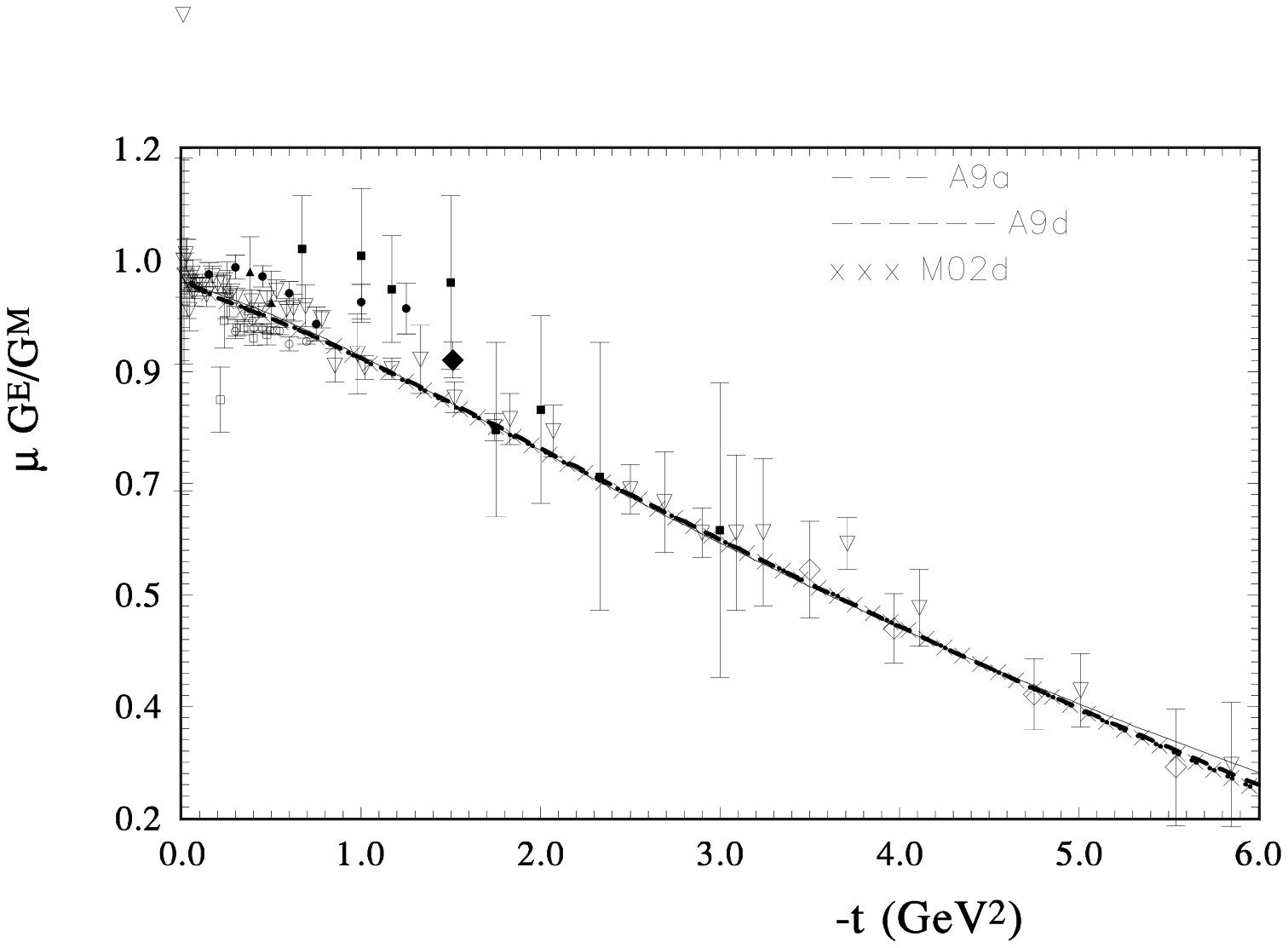}
\includegraphics[width=0.4\textwidth] {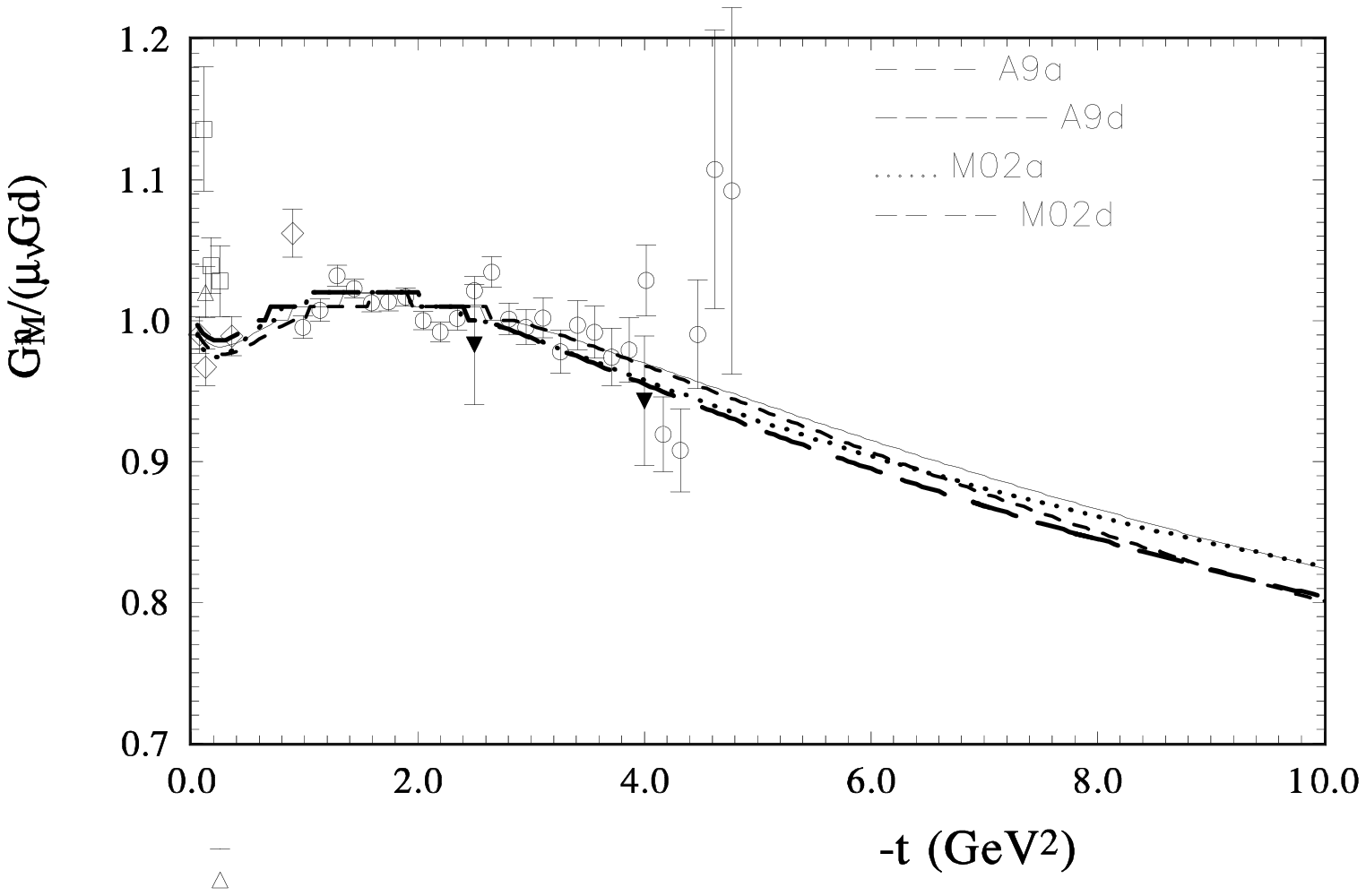}
\end{center}
 \caption{ The model description of the electromagnetic form factors for the proton(left)
    $\mu G_{E}^{p}/G_{M}^{p}$ and
  the neutron (right) $G_{m}^{n}/(\mu Gd)$ with the different PDFs.
 } \label{Fig_2}
\end{figure}

 %\section{Flower dependence of GPDs }

   Now let us examine  separate contributions of the $u$ and $d$ quarks to the
   electromagnetic form factors in our model of the $t$-dependence of  GPDs.
   We take  PDFs of  \cite{ABKM09} which give the one of the best descriptions of the electromagnetic form factors. We analyze the two cases: first - the base variant of  GPDs with only 4 free variation parameters,
  second - with the maximum number of  free variation parameters - $10$.
  %In last case some parameters reflect the flavor dependence of the GPDs.
\ba
{\cal{H}}^{q} (x,t) \  =&& q(x)_{u} \   exp [  \alpha  \ (a_{5} x(1-x)+
 (1-x)^{a_{1}} / (\epsilon +x)^{a_{2} }  \ t ]  \\ + && q(x)_{d} \   exp [  \alpha a_{3} \ (a_{6} x(1-x)+
 (1-x)^{a_{1} a_{4}} / (\epsilon +x)^{a_{2} }  \ t ]. \\ \nonumber
%\ea
%\ba
{\cal{E}}^{q} (x,t) \  = && q(x)_{u}(1-x)^{k_{1}} \   exp [  \alpha  \ (a_{5} x(1-x)+
 (1-x)^{a_{1}} / (\epsilon +x)^{a_{2} }  \ t ]   \\ \nonumber
+ && q(x)_{d}(1-x)^{k_{2}} \   exp [  \alpha a_{3} \ (a_{6} x(1-x)+
 (1-x)^{a_{1} a_{4}} / (\epsilon +x)^{a_{2} }  \ t ].   \nonumber \ea
 Here the parameters $a_{3},a_{4},a_{5},a_{6}$ represent the flavor dependence of the Regge part of  GPDs
  and the parameters   $k_{1},k_{2}$ are responsible for the  flavor dependence of the spin-dependent part of  PDFs.
  If we take the PDFs sets from  \cite{ABKM09} we obtain the small difference in  $\sum \chi^2$ in the descriptions
  of the electromagnetic form factors in these two cases, only $25 \% $. However, the number of  free parameters differs essentially: $4$ and $10$.
  Further increase in the number of  free parameters leads to  a very small decrease in  $\sum \chi^2$.

  %==================Figs.2 =============================
\label{sec:figures}
\begin{figure}
\vspace{-1cm}
\begin{center}
\includegraphics[width=0.4\textwidth] {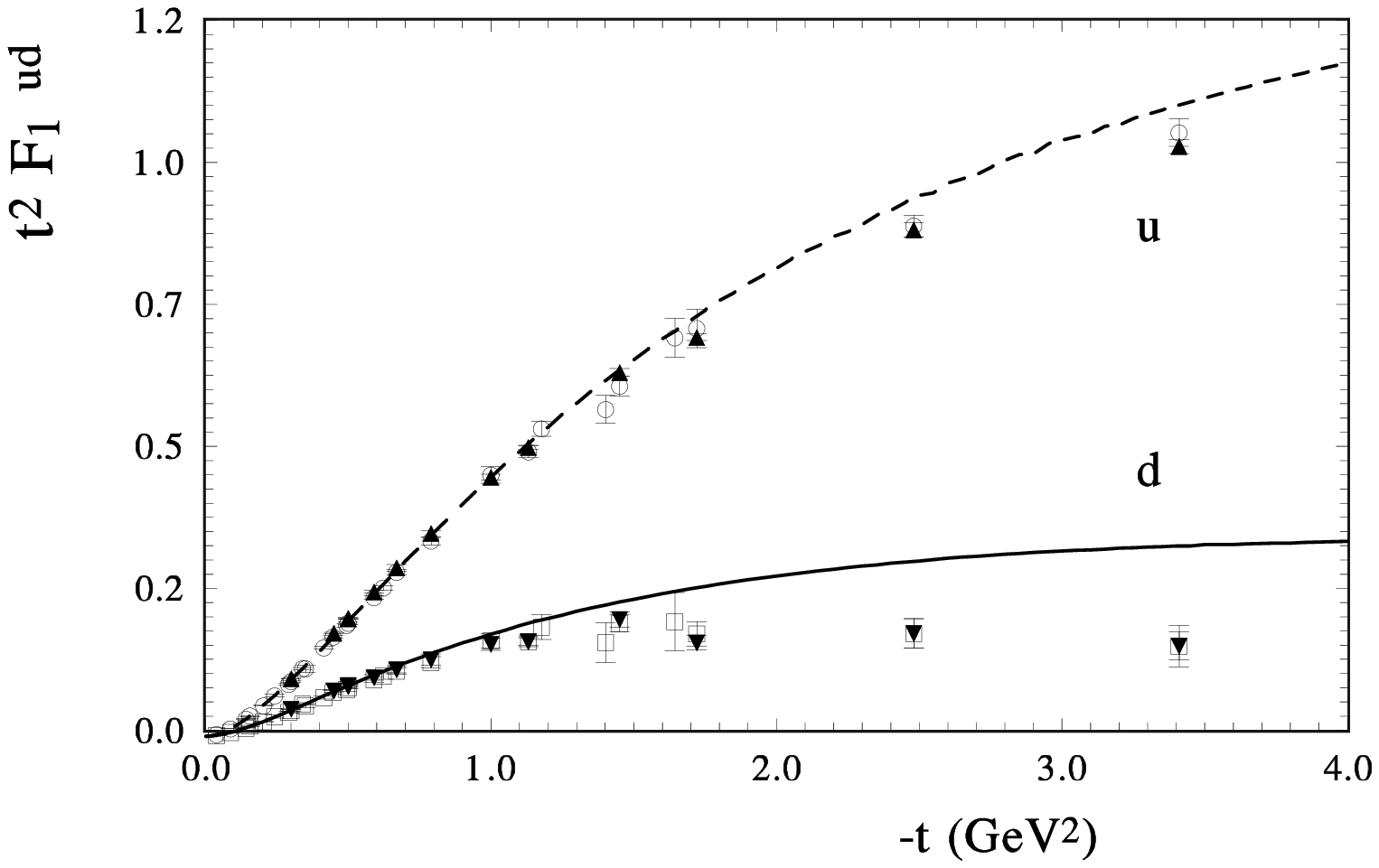}
\includegraphics[width=0.4\textwidth] {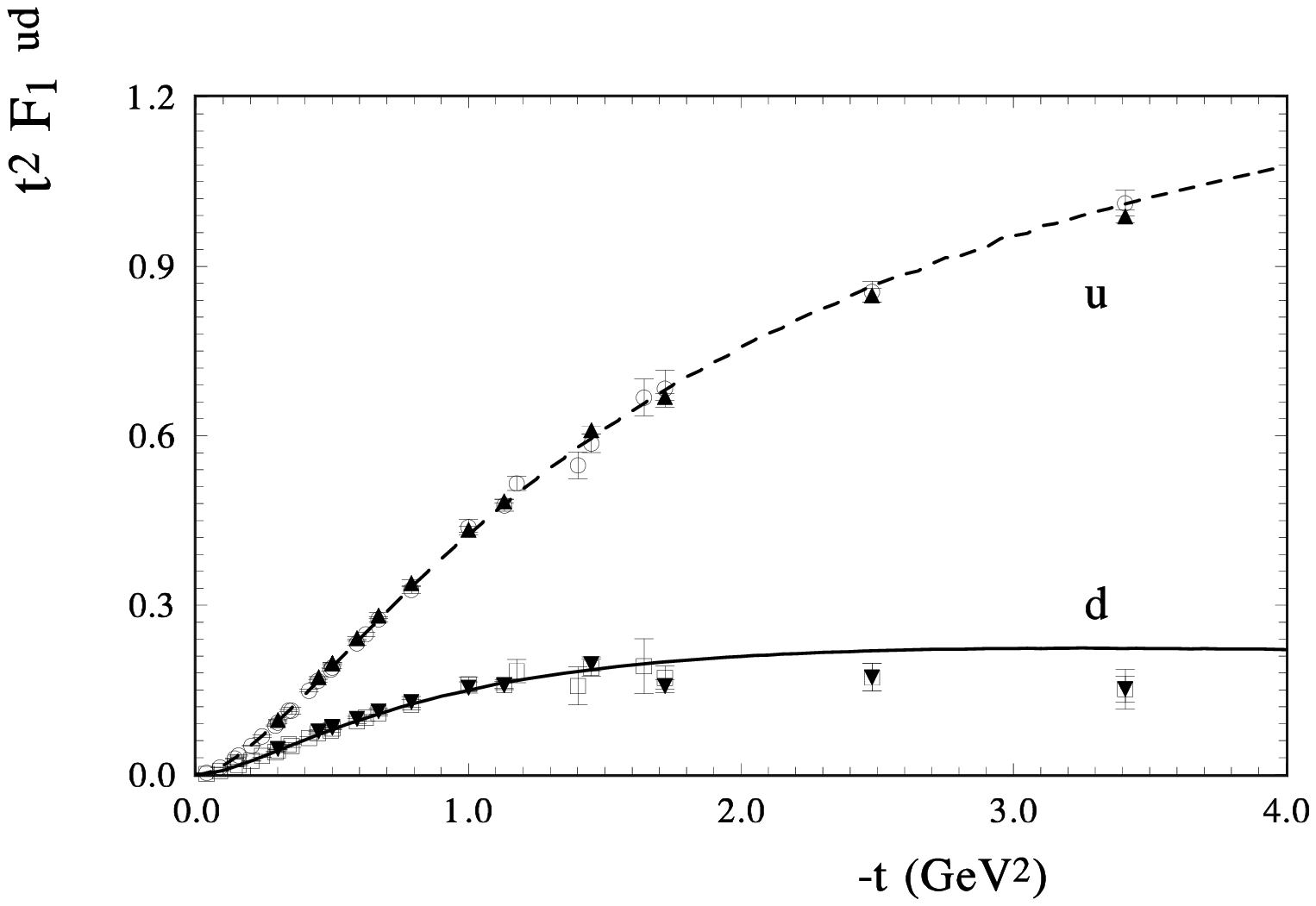}
\end{center}
 \caption{ The $u$ and $d$ quarks contributions to the $t^2 \ F_{1}(t)$:
    - the fit with $4$ free parameters (left) and  with $10$ free parameters (right).
     The data take from \cite{Rio2}.
 } \label{Fig_2}
\end{figure}
    %==================Figs.3 =============================
\label{sec:figures}
\begin{figure}
\vspace{-1cm}
\begin{center}
\includegraphics[width=0.4\textwidth] {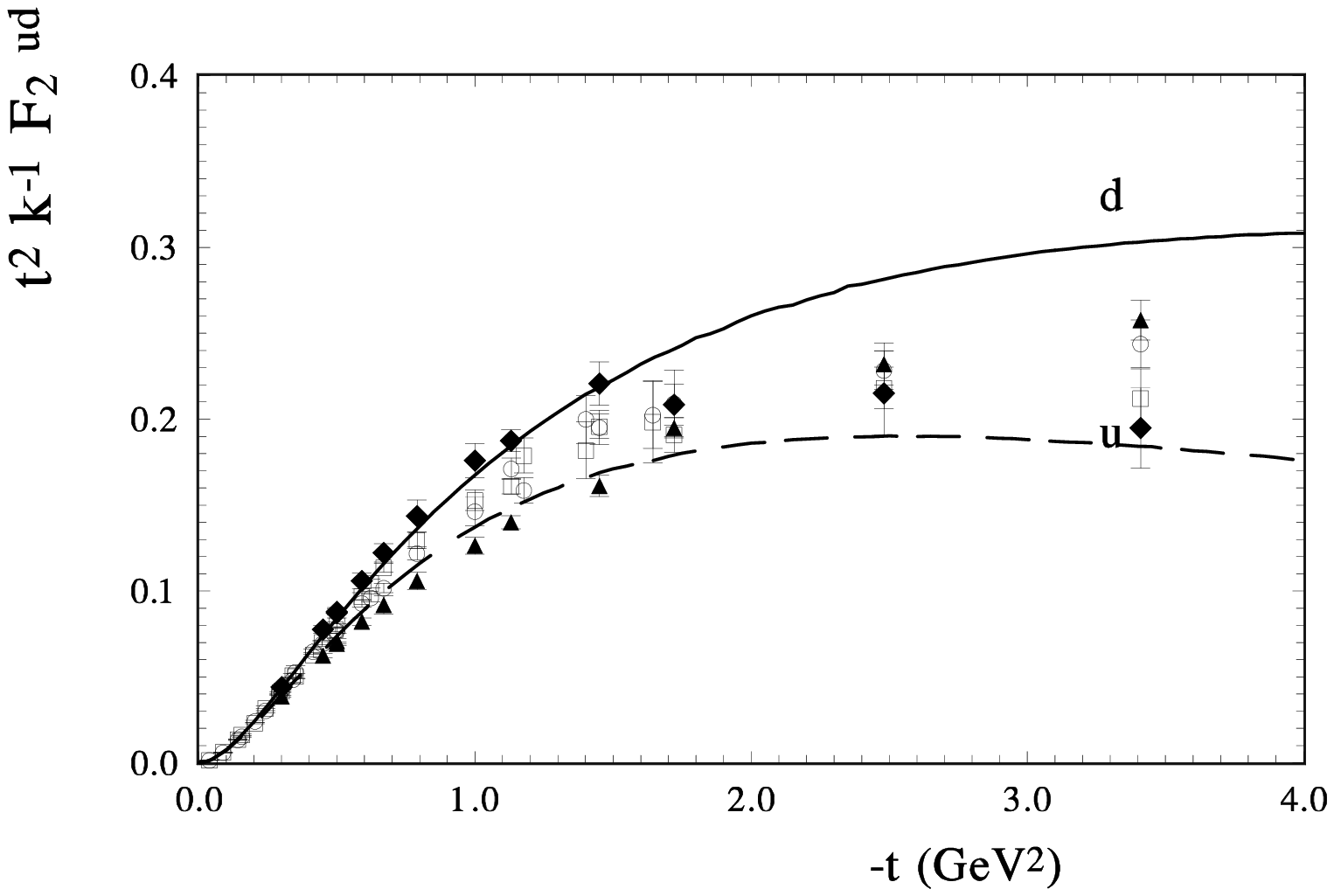}
\includegraphics[width=0.4\textwidth] {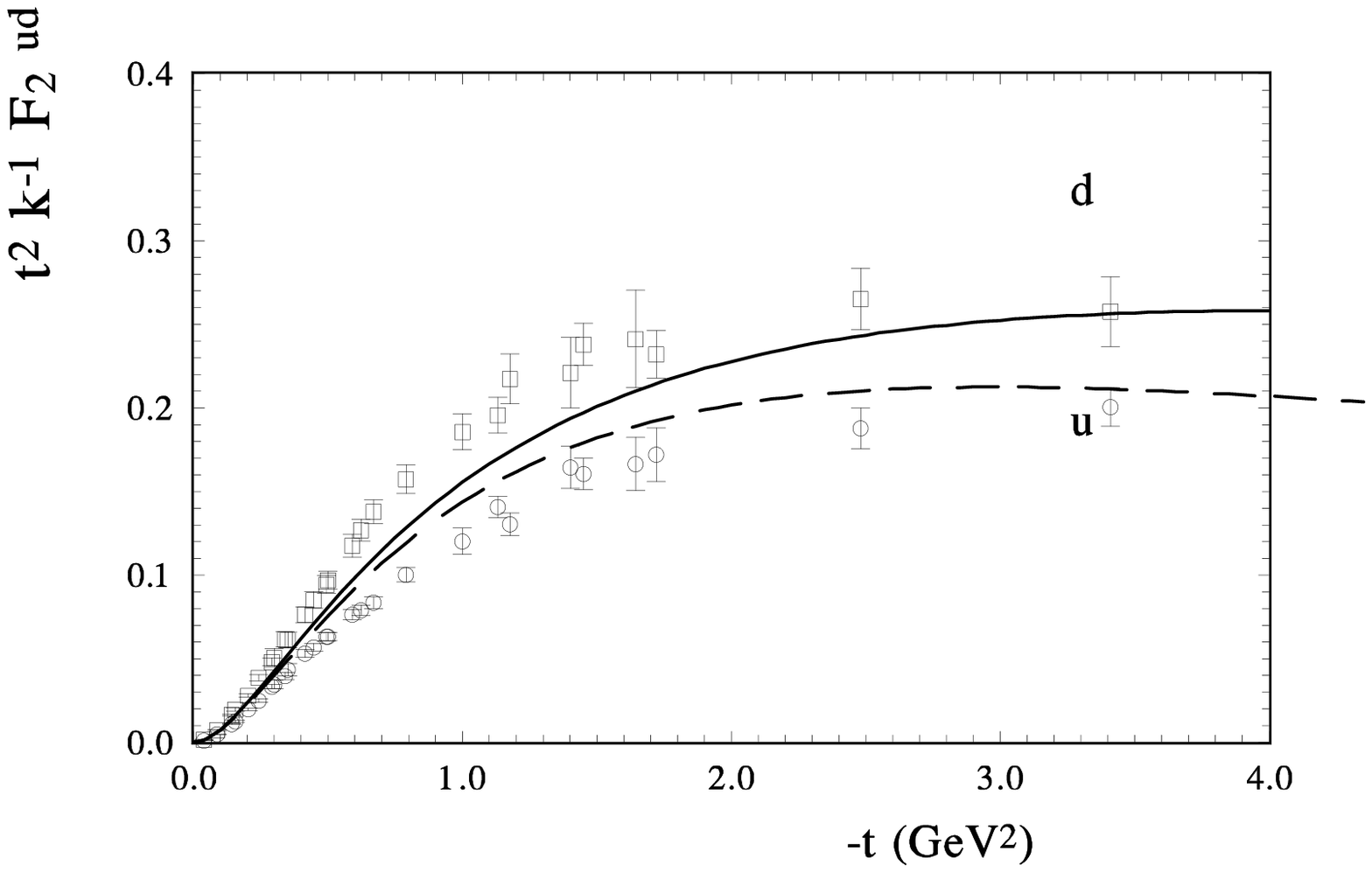}
\end{center}
 \caption{ The same as in Fig.2 for the $k^{-1} t^2 \ F_{2}(t)$.
 } \label{Fig_2}
\end{figure}

  The $u$ and $d$ quark contributions to  $F_{1}(t)$ multiplied by $t^2$ is shown in Fig.2.
    We compare  the fits with $4$ free parameters (left) and   $10$ free parameters (right).
    It is clear that the difference is very small. Only the $d$ quark contribution is slightly less
    in the last case. However, the $t$-dependence in  both the  cases is practically the same.
    In Fig.2, we present the same calculations for  $F_{2}(t)$. Again, the contribution of
    the $d$ quark decreases in the case of a large number of  free parameters.
    % with heavily  flavor dependence.
 %   But in both cases the contribution of the $d$ quarks exceeds the
 %   contribution of the $u$ quark. 
   Despite the large number of the free parameters,
    our calculations better coincide with  extractions of the $u$ and $d$ quark
    contributions up to $-t=2 $ GeV$^2$   \cite{Rio2}.
  The $u$ and $d$ quark contributions to $F_{1}(t)$ (left) and $F_{2}(t)$ (right)
  at large momentum transfer are shown in Fig.3. It is clear that at large $t$ the behavior of the
  $u$ and $d$ quark contributions is the same.
 % And we see that the $d$ contribution to  $F_{2}(t)$
 %  exceeded the $u$ contribution.

%\section{Flavor contributions at large momentum transfer}

%==================Figs.4 =============================
\label{sec:figures}
\begin{figure}
\vspace{-1cm}
\begin{center}
\includegraphics[width=0.4\textwidth] {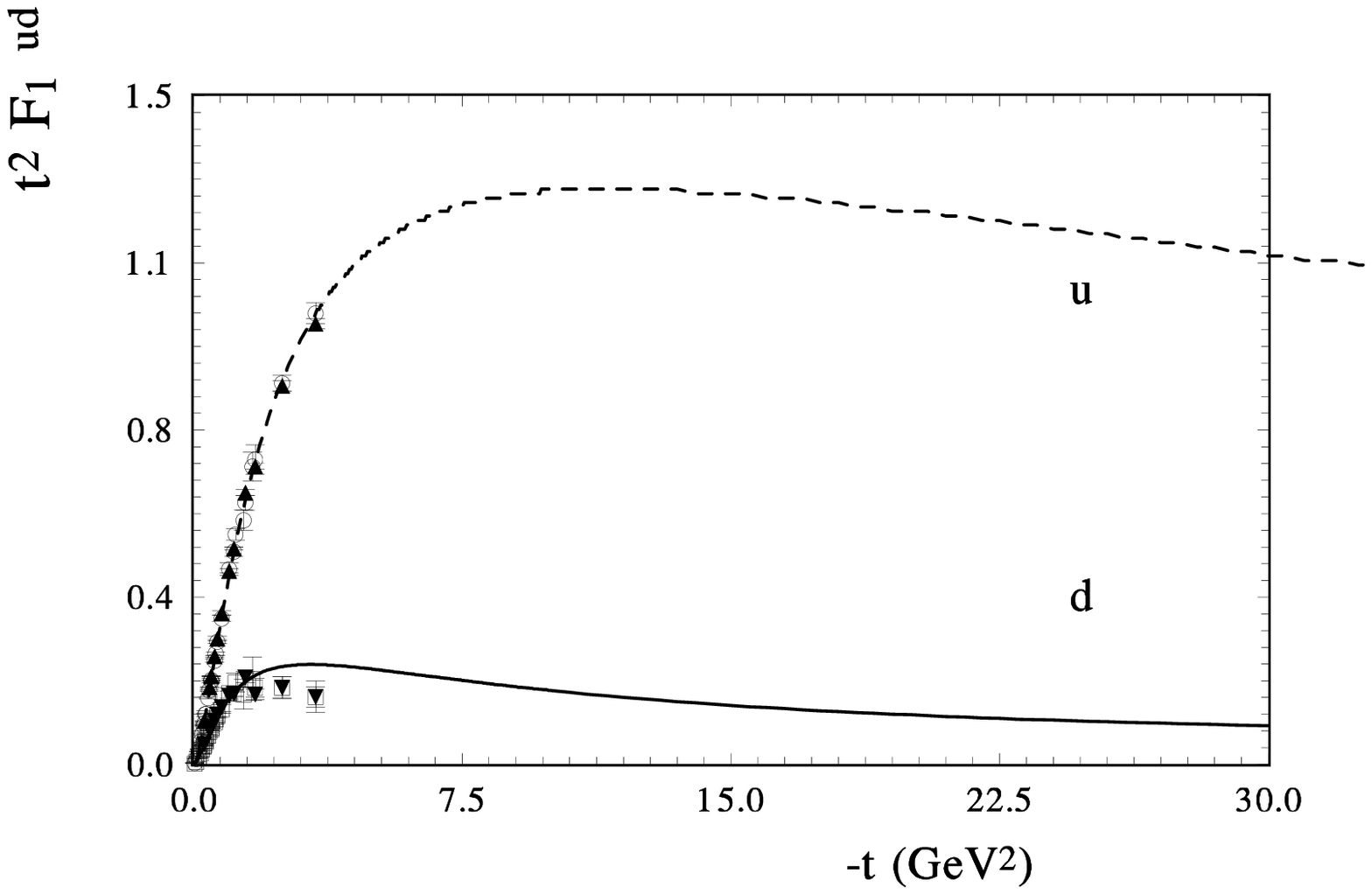}
\includegraphics[width=0.4\textwidth] {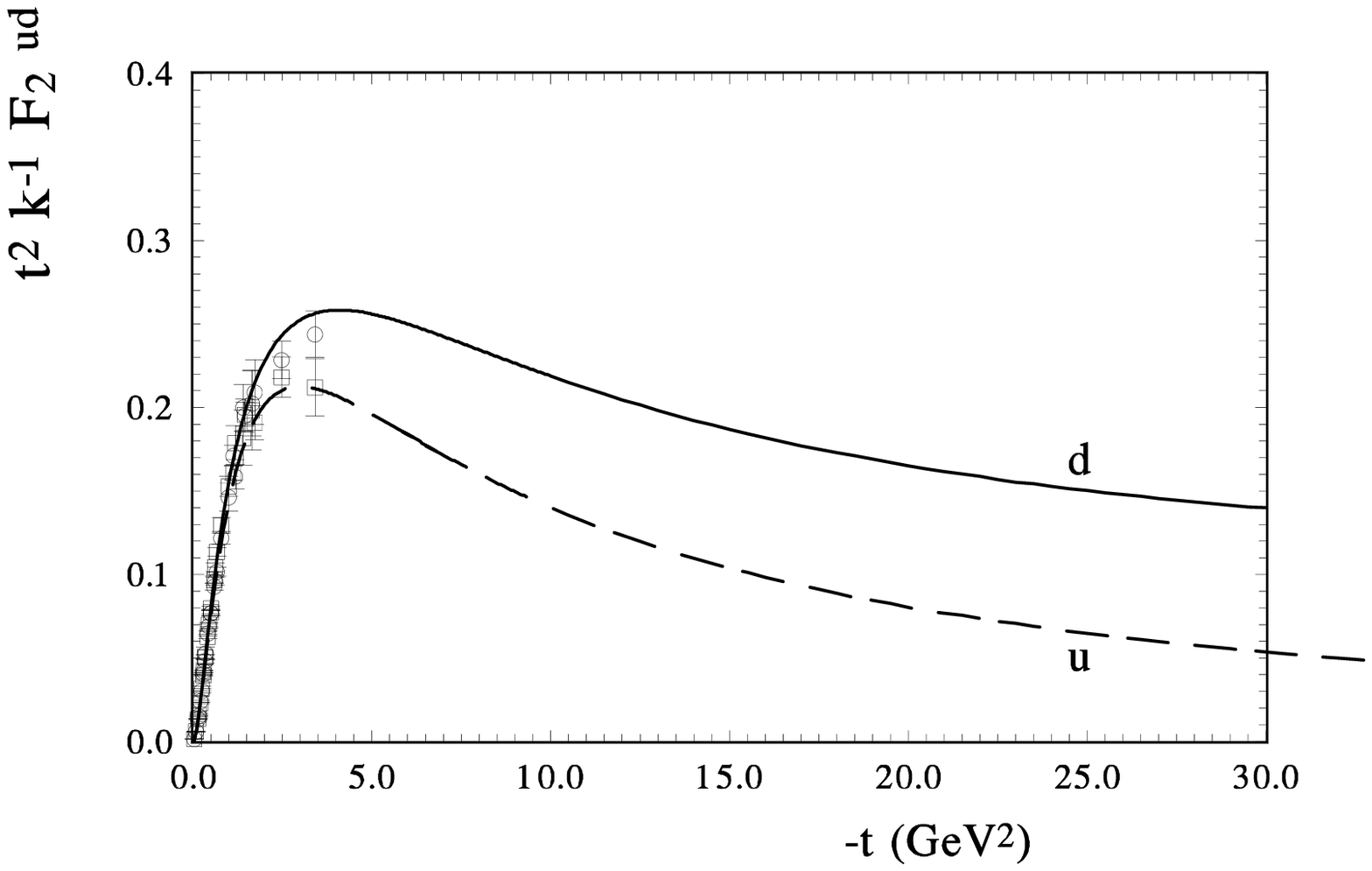}
\end{center}
 \caption{The $u$ and $d$ quarks contributions in $F_{1}(t)$ (left) and in $F_{2}(t)$ (right)
  at large momentum transfer.
 } \label{Fig_2}
\end{figure}

%\section{Conclusion }
 Our analysis of PDFs sets of the different Collaborations show a large difference
    in the descriptions of the electromagnetic form factors of the proton and neutron.
    The best result can be obtained with PDFs sets of  \cite{ABKM09} and
     \cite{ABM12}. These sets lead to minimum of  $\sum \chi^2$.
    They also show the small dependence of the  GPDs on the increasing  different
    free parameters. The obtained $t$ dependence of  GPDs has a simple form and
    a small number of the free parameters.

        The flavor dependence in these cases in most part comes from the spin dependent
        part of  PDFs.  We obtained the good descriptions of the electric and magnetic form factors
        of the proton and neutron simultaneously. We found that  different PDFs
        gave almost the same descriptions of the proton form factors at small momentum transfer.
        The difference appear only at large $t$. Our calculations of the $u$ and $d$ quark contributions show the same $t$ dependence
        at large $t$. % contrary to the results of the  \cite{Liuti2}.
      %  The $d$ contribution exceed the $u$ contribution in the $F_{2}$ form factors.

%\begin{footnotesize}
%\bibliography{abbr_long,pubext}

\begin{thebibliography}{99}
%\renewcommand{\baselinestretch}{0.5}
%\renewcommand{\baselinestretch}{0.5}
%\linespread{.5}
%\begin{spacing}{0.9}





%\bibliofont{ \rm}
%\begin{small}
%{\small
% \bibitem{DMuller94} D. Muller, D. Robaschik, B. Geyer, F.M. Dittes and J. Horejsi, Fortsch. Phys. %{\bf 42}, (1994) 101;

%1x
 \bibitem{Ji97} X.D. Ji, Phys. Lett. {\bf 78} , (1997) 610; Phys. Rev D {\bf 55} (1997) 7114; D. Muller {\it et al.}, Fortsch. Phys. {\bf 42}, (1994) 101;
      Radyushkin, A.V.,  Phys. Rev. D {\bf 56},  5524 (1997).
% \bibitem{R97}  Radyushkin, A.V.,  Phys. Rev. D {\bf 56},  5524 (1997)  .

\bibitem{Liuti1} G.R. Goldstein, J.O. Hernandez, S. Liuti,  Phys.Rev. {\bf D84} 034007 (2011).
% arXiv:1206.1876 v3.

\bibitem{Liuti2}J.O. Gonsales-Hernandes  {\it et al.}, arXiv:1206.1876 v3.

%
 \bibitem{Kroll04}  M.Diehl {\it et al.},  Eur.Phys. J. C  {\bf 39} (2005) 1.
%% arXiv [hep-ph/0408173].

\bibitem{Yuan03} F. Yuan, Phys. Rev. D, {\bf 69}, 051501(R) (2004) .
%% arXiv [hep-ph/0311288].

\bibitem{ST-PRDGPD}  O. Selyugin, O. Teryaev, Phys. Rev.
  {\bf D 79} 033003 (2009); % arXiv [0901.1786]:


 \bibitem{MRST02}  A.D. Martin  {\it et al.}, Phys. Lett. B  {\bf 531} (2002) 216.


\bibitem{R04} M. Guidal, % M.V. Polyakov,  A.V.  Radyushkin,  and M. Vanderhaeghen,
   {\it et al.}, Phys. Rev. D {\bf 72 }, 054013 (2005) .

\bibitem{Sel-Sp12}  O. Selyugin
%The structure of nucleons and the description of the electromagnetic form factors
 Intern. Simposium "SPIN in High Energy Physics", Dubna, (2012).
 %   arXiv:1304.2127


      \bibitem{ABKM09}S. Alekhin {\it et al.}, Phys.Rev. {\bf D81}
, 014032 (2010); % arXiv:0908.2766. %    \bibitem{ABM11}
 \bibitem{ABM12}  S. Alekhin, J. Blu"mlein, and S. Moch, Phys.Rev. D86
, 054009 (2012). % arXiv:1202.2281.

          \bibitem{Kh12}  H. Khanpour {\it et al.}, arXiv:1205.5194


       \bibitem{Rio2} G.D. Gates {\it et al.}, Phys.Rev.Lett. {\bf 106} 252003 (2011);
       I.A. Qattan and J. Arrington, Phys.Rev. {\bf C86} 065210 ( 2012)



 %                HERAPDF1.5 [18, 19],
%  \bibitem{HERAPDF1.5}

%                 JR09 [20, 21],
 %  \bibitem{JR09} P. Jimenez-Delgado and E. Reya, Phys.Rev.D80, 114011 (2009), arXiv:0909.1711


 %                 MSTW    [22] and
  % \bibitem{MSTW}  A. Martin, W. Stirling, R. Thorne, and G. Watt, Eur.Phys .J. C63, 189 (2009), % %
   %  arXiv:0901.0002.

 %                   NN21 [23],
    %    \bibitem{NN21} R. D. Ball et al. , Nucl.Phys. B855 , 153 (2012), arXiv:1107.2652.


   %while
   %              CT10 [24] s
  % \bibitem{CT10} H.-L. Lai et al., Phys.Rev. D82, 074024 (2010), arXiv:1007.2241.


 %           [25]
   % \bibitem{BD10}        J. Baglio and A. Djouadi, JHEP 1010, 064 (2010), arXiv:1003.4266

%[1] S. Alekhin, Phys.Lett. B519, 57 (2001), hep-ph/0107197.
%[2] S. Moch, J. Vermaseren, and A. Vogt, Nucl.Phys.B688, 101 (2004), hep-ph/0403192.
%[3] A. Vogt, S. Moch, and J. Vermaseren, Nucl.Phys.
%           B691, 129 (2004), hep-ph/0404111.
%[4] W. van Neerven and E. Zijlstra, Phys.Lett.B272, 127 (1991).
%[5] E. Zijlstra and W. van Neerven, Phys.Lett.B273, 476 (1991).
%[6] E. Zijlstra and W. van Neerven, Nucl.Phys.B383, 525 (1992).
%[7] E. Zijlstra and W. van Neerven, Phys.Lett.B297, 377 (1992).
%[8] S. Moch and J. Vermaseren, Nucl.Phys.B573, 853 (2000), hep-ph/9912355.
%[9] S. Moch, J. Vermaseren, and A. Vogt, Phys.Lett.B606, 123 (2005), hep-ph/0411112.
%[10] J. Vermaseren, A. Vogt, and S. Moch, Nucl.Phys.B724, 3 (2005), hep-ph/0504242.
%[11] R. Hamberg, W. van Neerven, and T. Matsuura, Nucl.Phys.
%              B359, 343 (1991).
%[12] R. V. Harlander and W. B. Kilgore, Phys.Rev.Lett.88, 201801 (2002), hep-ph/0201206.
%[13] C. Anastasiou, L. J. Dixon, K. Melnikov, and F. Petriell
%           o, Phys.Rev.Lett.91, 182002 (2003), hep-ph/0306192.
%[14] C. Anastasiou, L. J. Dixon, K. Melnikov, and F. Petriell
%           o, Phys.Rev.D69, 094008 (2004), hep-ph/0312266.
%[15] S. Catani, L. Cieri, G. Ferrera, D. de Florian, and M. Grazzini, Phys.Rev.Lett.103, 082001 (2009),arXiv:0903.2120.
 % \bibitem{MRST02}  A.D. Martin  {\it et al.}, Phys. Lett. B  {\bf 531} (2002) 216.
%
%
%%
%9
%
%\bibitem{Ernst60} F.J. Ernst, R.G. Sachs, and K.C. Wali Phys. Rev. {\bf 119} (1960) 1105.
%10
%\bibitem{Sachs62}R.G. Sachs, Phys.Rev. {\bf 126} (1962) 2256.
%\bibitem{Rosenbluth} M.N. Rosenbluth, Phys.Rev.{\bf 79}(1950) 615;
%\bibitem{Akhiezer} A.I. Akhiezer and M.P. Rekalo Sov.J.Pat.Nucl. {\bf 3} (1974) 277.


%13
%\bibitem{Arnold} R.G. Arnold, C.E. Carlson, and F. Gross,
 %            Phys.Rev. C {\bf 23} (1981)  363.
%14
%bibitem{Guichon} P.A.M. Guichon, M. Vanderhaeghen,
% Phys.Rev,Lett. {\bf 91}(2003) 142303-1;
% P.G. Blunden, W. Melnitchouk, A. Tjon,
% Phys.Rev,Lett. {\bf 91}, 142304-1 (2003);
%  Chen Y.C., {\it et al.}, Phys.Rev.Lett. {\bf 93}, 122301-1 (2004);
% M. P. Recalo, E. Tomasi-Gustafssn, Eur.Phys.J. {\bf A22} (204) 331;
% S. Dubnichka, E. Kuraev, M. Secansky, A. Vinnikov, hep-ph/0507242.
%
%bibitem{Qat05} I. A. Qattan,  {\it et al.}, Phys.Rev. Lett.  {\bf 94} (2005) 142301.
%  nucl-ex/0410001.



%%20
%\bibitem{Collins97} J. Collins, L. Frankfurt, and  Strikman M., Phys. Rev. D {\bf 56} (1997) 2982.
%\bibitem{R98} A. V. Radyushkin, Phys. Rev. D {\bf 58} 114008 (1998)  [arXiv:hep-ph/9803316].
%%22
%\bibitem{Burk00} M. Burkardt, Phys. Rev. D {\bf 62} 071503(R) (2000) .
%
%
% \bibitem{Brodsky81} S.J. Brodsky, T. Huang and G.P. Lepage, in Particles and Fields 2, Proceedings of the Banff Summer Institute, Bunff, Alberta, 1981, edited by A.Z. Capri and A.N. Kamal (Plenum, NY, 1983) 143.

%\bibitem{Goeke01} K. Goeke, M.V. Polyakov, M. Vanderhaeghen,  Prog.Part.Nucl.Phys.{\bf 47} (2001) 401.
% % [arXiv:hep-ph/0106012].
%%25

%  %    [arXiv:hep-ph/0410251].
%
%\bibitem{Stol01} P. Stoler,  Phys.Rev., D {\bf 65} (2002) 053013.
%  % hep-ph/0108257.
%\bibitem{Stol02} P. Stoler,  Phys.Rev.  Lett.,  {\bf 91} (2003) 172303.
 % hep-ph/0210184.


%
%%30
%\bibitem{Burk04} Burkardt M., Phys.Lett. B {\bf 595}, 245 (2004) .
%% arXiv [ hep-ph/0401159].



%%\cite{Brash:2001qq}
%\bibitem{Brash:2001qq}
%  E.~J.~Brash, A.~Kozlov, S.~Li and G.~M.~Huber,
%  %``New empirical fits to the proton electromagnetic form factors,''
%  Phys.\ Rev.\  C {\bf 65}, 051001 (2002)
%  [arXiv:hep-ex/0111038].
%  %%CITATION = PHRVA,C65,051001;%%
%
%



%35
%\bibitem{Jones00} M.K. Jones {\it et al.}, Phys.Rev.  Lett.,  {\bf 84} (2000) 1398.
%
%\bibitem{Gayou01} O. Gayou {\it et al.}, Phys.Rev.  C  {\bf 64} (2001) 038202.
%\bibitem{Gayou02} O. Gayou {\it et al.}, Phys.Rev.  Lett.,  {\bf 88} (2002) 092301.
%\bibitem{Sill93} A.F. Sill  {\it et al.}, Phys.Rev.  D  {\bf 48} (1993) 29.
%\bibitem{Brodsky03}  S.J.  Brodsky ,  hep-ph/0208158 .
%
%%40
%\bibitem{Punjabi05} V. Punjabil  {\it et al.}, Phys.Rev.  C  {\bf 71} (2005) 055202.

%\bibitem{Arr05}  J.Arrington ,   Phys.Rev.  C  {\bf 71 } (2005) 015202.
%% arXiv [hep-ph/0408261] .
%
%\bibitem{Hu06} B. Hu {\it et al.} ,  nucl-ex/0601025.
%
%
%\bibitem{Rock82} S. Rock {\it et al.}, Phys.Rev. Lett.  {\bf 49} (1982) 1139.
%%   Phys. Rev. {\bf 59} (1999) 105006.
%  %hep-ph/981135

%\bibitem{Graw07}  C.B. Grawford,  {\it et al.}, Phys.Rev. Lett.  {\bf 98} (2007) 052301.
% % nucl-ex/00609007.
%
%\bibitem{Glazier04} D.I. Glazier  {\it et al.}, Eur.Phys.J.  A  {\bf 24 } (2005) 101.
%%     arXiv[nucl-ex/0410026.
%
%\bibitem{Plaster05} B. Plaster  {\it et al.}, Phys.Rev.  C  {\bf 73} (2006) 025205.
%
%\bibitem{Kubon02}  G. Kubon,  {\it et al.}, Phys. Lett. B  {\bf 524} (2002) 26.
%  nucl-ex/0107016.

%\bibitem{Madey03}  R. Madey,  {\it et al.}, Phys.Rev. Lett.  {\bf 91} (2003) 122002.
%%  nucl-ex/0308007.
%
%\bibitem{Warren04} I. G. Warren,  {\it et al.}, Phys.Rev. Lett.  {\bf 92} (2004) 042301.
%%  nucl-ex/0410001.
%%50
%\bibitem{Anderson07} B. Anderson,  {\it et al.}, Phys.Rev.  C  {\bf 75} (2007) 034003.
%% nucl-ex/0605006.
%

%\bibitem{Rock82} S. Rock,  {\it et al.}, Phys.Rev. Lett.  {\bf 49} (1982) 1139.
%  nucl-ex/0410001.
%\bibitem{Qat06} I. A. Qattan, Ph.D. Thesis, Northwestern University, Advisor: Ralph E. Segel; nucl-ex/0610006.
%\bibitem{Ji96}  X.D. Ji,  in Proceed. the 12th International Symposium on High-Energy Spin Physics, Amsterdam, Sept., 1996.
%    %    hep-ph/9610369
%
%\bibitem{Ji97b} A.V. Belitsky, X.D. Ji, Phys. Lett.  B {\bf 538} , (2002) 289.
%% hep-ph/0203276
%% Phys. Rev D {\bf 55} (1997) 7114.
%
%% \bibitem{Brodsky00} S.J. Brodsky,  Dae Sung Hwang, Bo-Qiang Ma, and  Ivan Schmidt,
% Nucl.Phys. B {\bf 578} (200) 326.
% hep=ph/0003082.

%\bibitem{ter1}  O.V. Teryaev,  hep-ph/9904376



%\bibitem{RayT} H. Dahiya, A. Mukherjee, S. Ray, [hep-ph]/0705.3580.
%
%\bibitem{Vanderh} C.E. Carlson and M. Vanderhaeghen,
%             Phys.Rev.Lett., {\bf 100} (2008) 032004.
%
%\bibitem{BurkZ} M. Burkardt, [hep-ph]/0709.2966v2(October, 2007).
%
%\bibitem{MillerZ} G. A. Miller,
%  Phys.Rev.Lett., {\bf 99} (2007) 112001;
%[hep-ph]/0705.2409v2(May, 2007); [nucl-th]/0802.2563v2(20 Feb. 2008).
%%60
%\bibitem{Arrington-ron} Kelly J.J., Phys. Rev. C, {\bf  66}, 065203 (2002) .
%
%%\bibitem{Arrington-ron} Arrington
%
%
%%62
%\bibitem{Gockler04} M. Goeckeler, {\it et al.}, Nucl.Phys. (Proc.Suppl.) {\bf 128}, 203 (2004).
% \bibitem{Hagler05} Ph. Hagler, {\it et al.}, Eur.Phys. J., A {\bf 24}, 29 (2005).

%}
%\end{small}


\end{thebibliography}

 {\footnotesize
\begin{spacing}{0.5}
%{\small

\end{spacing}

}
%\end{footnotesize}

\end{document}